\documentclass[aps,pra,reprint,superscriptaddress,amsmath,amssymb]{revtex4-2}

\usepackage{hyperref}
\usepackage{physics}
\usepackage{pgfplotstable}
\usepackage{xr}
\usepackage{microtype}
\usepackage{verbatim}

\pgfplotsset{compat=1.18}

\graphicspath{{./img/}}

\newcommand{\kon}{k_\textrm{on}}
\newcommand{\koff}{k_\textrm{off}}

\newcommand{\kopti}{k_\textrm{off}^*}

\newcommand{\cDNA} {c_s}
\newcommand{\cbulk} {c_\textrm{bulk}}

\newcommand{\ctarget} {c_\textrm{target}}

\newcommand{\cDNAbar} {\bar c_s}

\newcommand{\cbulkbar} {\bar c_\textrm{bulk}}
\newcommand{\ctargetbar} {\bar c_\textrm{target}}

\newcommand{\Nstates}{N_\textrm{b}}

\newcommand{\Jinf}{J_\infty}

\definecolor{Red}{RGB}{220,35,10}
\definecolor{Blue}{RGB}{15,75,210}
\definecolor{Green}{RGB}{5,180,55}

\begin{document}

    \title{Target search on networks-within-networks with applications to protein-DNA interactions}
    
    \author{Lucas Hedström}
    \email[Corresponding author: ]{lucas.hedstrom@umu.se}
    \author{Seong-Gyu Yang}
    \author{Ludvig Lizana}
    \affiliation{Integrated Science Lab, Department of Physics, Ume\aa~University, Ume\aa, Sweden}
    
    \date{\today}

    \begin{abstract}
    We present a novel framework for understanding node target search in systems organized as hierarchical networks-within-networks. Our work generalizes traditional search models on complex networks, where the mean-first passage time is typically inversely proportional to the node degree. However, real-world search processes often span multiple network layers, such as moving from an external environment into a local network, and then navigating several internal states. This multilayered complexity appears in scenarios such as international travel networks, tracking email spammers, and the dynamics of protein-DNA interactions in cells. Our theory addresses these complex systems by modeling them as a three-layer multiplex network: an external source layer, an intermediate spatial layer, and an internal state layer. We derive general closed-form solutions for the steady-state flux through a target node, which serves as a proxy for inverse mean-first passage time. Our results reveal a universal relationship between search efficiency and network-specific parameters. This work extends the current understanding of multiplex networks by focusing on systems with hierarchically connected layers. Our findings have broad implications for fields ranging from epidemiology to cellular biology and provide a more comprehensive understanding of search dynamics in complex, multilayered environments.
    \end{abstract}
    
    \maketitle
    
    \section{Introduction}
    One of the most well-known results in node-search on complex networks is that the mean-first passage time (MFPT) is inversely proportional to the node degree~\cite{Noh2004RN}. However, in real-world scenarios, search processes typically occur on more complex structures with a networks-within-network-like organizations. In such cases, the node degree is but a crude proxy for search times. Imagine this example: say you wish to find a tourist site in a far-away city. First, you must travel through the airline transportation network to find the correct airport and then find a way through the local transportation system to reach the desired location. Next, you must navigate the local surroundings to find your intended endpoint. To estimate the total travel time, you need to consider the time spent on all these network layers, which is not simply proportional to the number of connections associated with the final destination.

This problem generalizes the standard node search on complex networks to systems with a networks-within-network organizations. Apart from travel, this class of search problems encompasses tracking down virus-infected computers or email spammers on Local Area Networks connected to the Internet~\cite{Shah2010_Virus}, finding super-spreaders in epidemic outbreaks (like COVID-19~\cite{Nielsen2021COVID19, Nielsen2023COVID19}), tracing contaminants in river networks or subway systems~\cite{Pinto2012diffusionsource}, or identifying leaders in organized crime or interconnected terrorist cells~\cite{Syed2020terror, Latora2004terror}.

\begin{figure}[h!]
    \includegraphics[width=0.9\columnwidth, page=1]{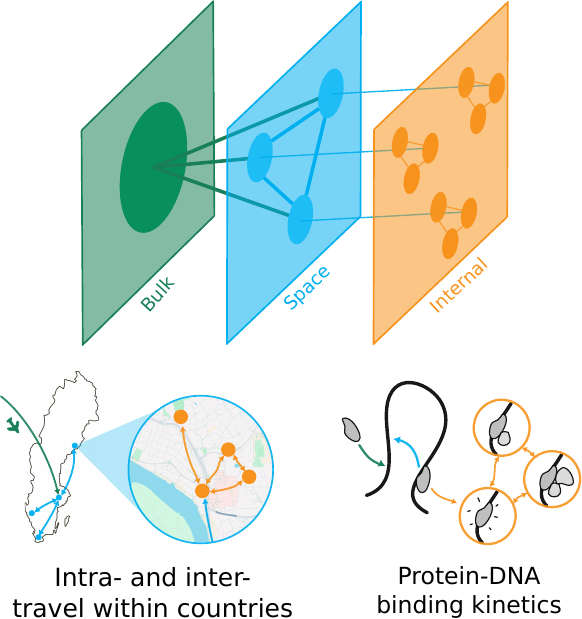}
    \caption{
    Our model considers a three-layer multiplex network.
    The top layer (green) represents inflow from outside, such as people traveling into a country or proteins diffusing onto DNA from the nucleoplasm.
    The middle layer (blue) reflects connections inside the system, like the national travel network or translocations between DNA segments.
    The bottom layer (yellow) captures internal degrees of freedom, such as city streets in a city or conformational states of protein complexes.
    }
    \label{fig:intro-schematic}
\end{figure}

These cases illustrate search on networks-within-networks in macroscopic scales. However, there are examples at microscopic scales, too. For example, critical cellular processes such as gene regulation, DNA repair, and transcription rely on specific protein complexes finding and binding designated target sites on DNA, typically a short base pair sequence. While searching, these complexes undergo switching between several internal states that regulate the sensitivity and speed of the search or direct the complexes to select parts of the genome. These internal states can represent conformational states or modifications by small molecules or protein subunits. In some cases, the switching is due to thermal fluctuations, where the searching complexes alternate between states having different free energies.
For example, the search-and-recognition mode of Transcription Factors (TF) that regulate genes~\cite{luking2022conformational} or sigma factors binding to RNA polymerases to shift the general gene expression of select gene families. Other cases are driven by non-equilibrium switching, hydrolyzing ATP molecules to release energy. One such case occurs in DNA repair where the MutS protein spends two ATP molecules when recognizing a mismatched base pair site and forms a sliding clamp conformation~\cite{acharya2003coordinated}.
This clamp is essential for recruiting the next protein in the repair cascade (e.g., MutL). 

However, describing the switching between the internal states is not enough to quantify or understand search times for targets on DNA. In addition to the proteins' internal dynamics, the search often occurs on complex media, like the nucleus or the chromosomes. These structures exhibit fractal-like properties often represented as complex networks~\cite{hedstrom2024identifying,lieberman2009comprehensive,kim2022fractal}.
Therefore, like navigating through a multi-layer transportation network when looking for the DNA-target search problem falls in the same class where proteins must navigate complex spatial structures in addition to being in the correct internal state to find and recognize designated targets.

Networks-within-networks, a specific subclass of multiplex or multilayer networks~\cite{Boccaletti2014Review_MN, Battiston2016MN, Mikko2014MN, Lee2015review_MN} that emphasize connections between distinct, often independent networks, whereas multiplex networks typically describe nodes participating in multiple networks simultaneously, fulfilling different roles on different layers.
As illustrated in Fig.~\ref{fig:intro-schematic}, we model a three-layer structure. The middle layer (blue) represents a spatial network, such as 3D contacts within a chromosome, where the target node resides. However, since the target can only be detected when the searcher is in the appropriate internal state, we associate each node in the middle layer with its own network of internal states (orange). We envision this spatial network in contact with an external source called the ``bulk'' (green).In protein target search, the bulk represents the cell volume surrounding DNA; in travel contexts, it could represent flights arriving from other continents.

In this paper, we develop a general theory for node target search on networks-within-networks. Specifically, the theory concerns the concentration of searchers (e.g., proteins) cascading through all network layers from the ``bulk'', where we are interested in the steady-state flux through a designated target node.
We find several closed-form expressions holding for general network configurations.
For example, we find that the steady-state flux $\Jinf$, our proxy for inverse MFPT, follows a universal relationship depending on two network-specific parameters $K$ and $\Gamma$ associated with the blue and orange layers, and the number of nodes $N$ in the spatial layer.
When the internal network has two nodes, like in a simple TF two-state target search~\cite{slutsky2004kinetics} we find the time spent on the network is given by a geometrical average of the on and off rates as $\sim (\kon \kon')^{1/2}$.
We also study the trade-off between having a large internal network yielding robustness to network perturbations but, at the same time, generally slowing down the search. 
    
    \section{Method}
\subsection{General theory} \label{subsection:general}

    \begin{figure}
        \includegraphics[width=1.0\columnwidth, page=2]{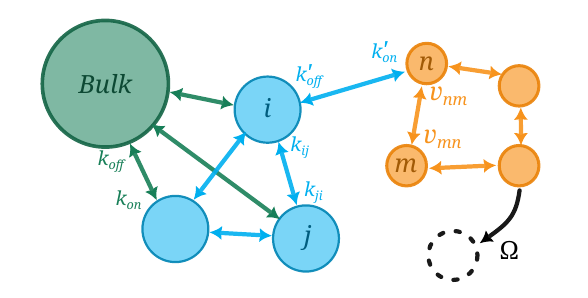}
        \caption{
        The three-layer multiplex network in Fig.~\ref{fig:intro-schematic} with rates. Every node in spatial space (blue) is connected to the singular bulk node with rates $\kon$ and $\koff$. The nodes in the spatial layer are numbered $i=1,\cdots, i,\cdots, N$ and connected by the rates $k_{ij}$. Every blue node has a network of internal states (orange) connected with rates $\kon'$ and $\koff'$. We denote the rates within the internal network by $\nu_{nm}$. For clarity, we only show one of the internal states. Only one of the internal networks has the target node, which has a connection to the absorbing node with an exit rate $\Omega$.
        }\label{fig:effective-model}
    \end{figure}
    
    The general networks-within-network search model is shown in Fig.~\ref{fig:effective-model}. the colors indicate three hierarchical layers: bulk (green), spatial (blue), and internal (orange). We envision the searcher as a walker, or a particle, that moves randomly on the spatial network from node $i$ to $j$ with a rate $k_{ji}$. At each node in the spatial network, the searcher can enter into a network of internal states, or internal degrees of freedom, with rate $\kon'$ via a few selected ``surface'' nodes $S$. Once inside the internal network, the searcher cannot leave a spatial node until it returns to one of the surface nodes. Within the internal network, the searcher changes its state or moves from node $n$ to $m$ with rate $\nu_{mn}$.

    The model also contains a source and a sink. These give rise to a cascading concentration flux through the network layers. The source comes from the bulk layer (green), representing a reservoir of searchers. A searcher may translocate from the bulk to the spatial network with rate $\kon$ and return with rate $\koff$. The sink is associated with an internal network node (orange) in a designated node in the spatial network (blue). This means that two conditions must be satisfied in order to find the target --- the searcher must be in at the right target node and in the correct internal state. If so, it can detect the target with rate $\Omega$ and get absorbed.
    
    To mathematically formulate our model, we write coupled equations for the concentration of searchers in each node at time $t$ as they move randomly through the network-within-network structure to find the target:
    \begin{align} \label{eq:EoM}
        \frac{d c_i}{dt} =  & \kon \cbulk -\koff c_i + \sum_j^N \left( k_{ij} c_j - k_{ji}c_i \right)\\\nonumber
        & + \sum_n^{N_b} \left(\koff' c_i^n - \kon' c_i \right)\delta_{n\in S}, \\\nonumber
        \frac{d c^n_i}{dt} = & \left(\kon' c_i - \koff' c^n_i \right)\delta_{n\in S} + \sum_m^{N_b} \left( \nu_{nm} c^m_i - \nu_{mn} c^n_i \right)\\\nonumber
        & - \Omega c^n_i \delta_{i,t} \delta_{n,t'}.
    \end{align}
    Here, $c_i$ ($i=1, 2,\cdots, N$) and $c_i^n$ ($n = 1, 2, \cdots, N_b$) denote the node concentration in the spatial and internal network, respectively. The first two terms of the upper equation describe the concentration fluxes from and to the bulk. The following two terms represent the random motion on the spatial network, and the last two terms indicate the flux between the spatial and the internal network. The notation $\delta_{n\in S}$ in the last two terms denotes we restrict the index to the subset $S$ of internal nodes that are connected to the spatial network (``surface'' nodes). 
    
    The lower equation follows a similar form as $c_i$. The first two terms describe the exchange between the spatial and internal network, and the next two terms represent the random motion over the internal network inside spatial node $i$. The last term is the sink term. It is proportional to the absorbing rate $\Omega$, where the Kronecker delta function ensures this term is zero unless the searcher is at the target node $i=t$ and in the correct internal state $n=t'$.
    
    One of the main goals of this paper is to better understand target search times. One common approach is to consider a single particle diffusing and then track how long it takes to reach the target, giving the first-passage time $\tau$~\cite{redner2001guide}. However, there is an alternative approach to finding $\tau$ that considers the constant flux through the target $\Jinf$. By assuming that, on average, only one protein molecule is present at any given time, the flux is identical to the constant rate of binding to the target, or the inverse of the MFPT~\cite{mcclintock1989noise,hu2006proteins,nyberg2021modeling,sokolov2005first}
    \begin{equation}
        \tau = \frac 1 \Jinf.
    \end{equation}
    We apply this approach here and calculate the steady-state flux as
    \begin{align} \label{eq:gen_J}
        J_\infty  = \Omega \ctargetbar,
    \end{align}
    where $\ctargetbar = \bar c_t^{t'}$ is the concentration at the target site at steady state.
    
\subsection{Simulations}

    To corroborate our analytical results, we performed stochastic simulations using the Gillespie algorithm. Each sample considers an individual searcher starting in a random node in the spatial network. Once the searcher finds the absorbing target, we record the search time and restart the simulation. If not otherwise specified, the rates were all set to one, except $\Omega = 10^4$, and a system size of $N = N_b = 10$.
    
    \section{Results}
\subsection{General expressions for target search}

    \begin{figure*}
        \centering
        \includegraphics[width=\textwidth]{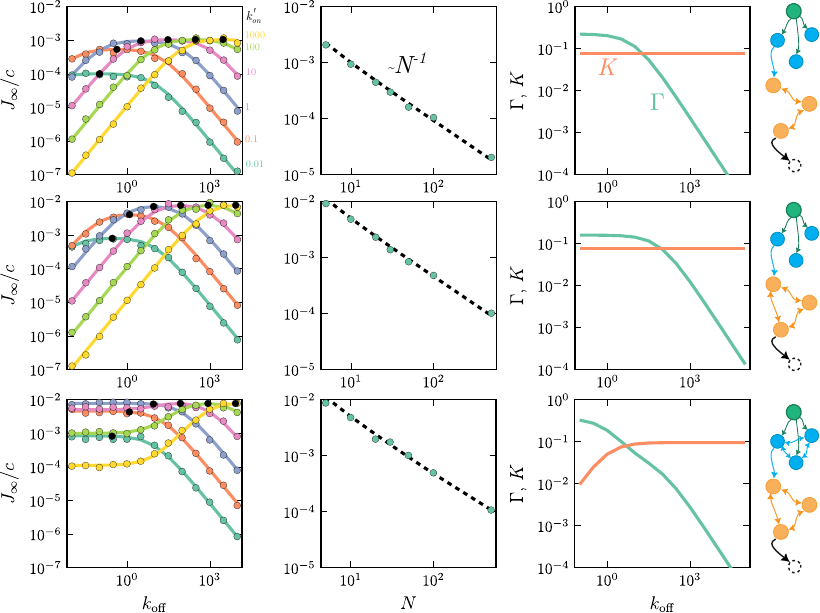}
        \caption{Optimal search times and size dependence for the linear chain (upper row) and random internal networks (middle row) with random spatial network (bottom row), showing both simulations (dots) and theoretical results (lines).
        \textbf{(left column)} We calculated the flux $\Jinf$ whilst varying the exit rate $\koff$ for different rates of switching to the internal state $\kon'$. In the two top cases, we see a distinct maximum for $\koff$ for different $\kon'$. At lower $\kon'$, the maximum is flanked by a plateau, which narrows for larger $\kon'$. With a spatial net, the maximum is less obvious except for larger $\kon'$.
        \textbf{(middle column)} The flux $\Jinf$ versus the system size $N$ using fix values of $\koff$ and $\kon'$. We see an inverse relationship between the quantities, $\Jinf \sim N^{-1}$.
        \textbf{(right column)} To better understand the results in the left column, we plot the parameters $K$ and $\Gamma$ normalized by their sum against $\koff$. For small $\koff$, we expect a linear increase in $\Jinf$. However, this is countered by $\Gamma$, which compared to the constant $K$ has a power-law decrease for larger $\koff$.
        }
        \label{fig:optimal_search}
    \end{figure*}
    
    In Sec.~\ref{subsection:general}, we gave the general expression for the target search flux in Eq.~\eqref{eq:gen_J}. Here, we derive an explicit formula that depends on the parameters associated with the spatial and internal networks by solving the general model in Eq.~\eqref{eq:EoM} in steady-state.
    
    First, we note that the total concentration can be written as a sum of three contributions
    \begin{equation} 
        c(t) = \cbulk(t) + \ctarget^\textrm{total}(t) + \cDNA^\textrm{total}(t),
    \end{equation}
    where $c_\textrm{target}^\textrm{total} = c_t(t) + \sum_{n=1}^{N_b} c_t^n(t)$ is the total concentration on the spatial target node $t$, and $c_s^\textrm{total}$ is the total concentration on the spatial network except on the spatial target node. That is,
    \begin{equation}
        \cDNA^\textrm{total}(t) = \sum_{i=1, i\neq t}^N \left(c_i(t) + \sum_{n=1}^{N_b} c_i^n(t)\right).
    \end{equation}
    However, since the node that contains the target inside its internal network is a small fraction of the spatial network ($N\gg 1$), we assume that
     \begin{equation} \label{eq:gen_total_c}
        c(t) \approx \cbulk(t) + \cDNA^\textrm{total}(t).
    \end{equation}
    
    In steady-state, all these concentrations are proportional to each other.
    This means we may write
    \begin{equation} \label{eq:gen_ct_cdna}
      \begin{split}
        &\ctargetbar = \Gamma \cbulkbar, \\
        &\cDNAbar^\textrm{total} = K (N-1)\frac{\kon}{\koff}\cbulkbar,
      \end{split}
    \end{equation}
    where the over-bar denotes the steady-state value, and $\Gamma$ and $K$ depend on the connectivity of the spatial and internal networks. In general, these terms are (See Appendix~\ref{appx:derive} for a detailed derivation)
    \begin{equation}\label{eq:factors}
        \begin{split}
    	\Gamma =& \frac{\kon}{\koff} \frac{\kon'}{\koff'}  \left( \sum_j^N W^{-1}_{t j} \right) \left( \sum_{m\in S} U^{-1}_{t' m} \right),\\
    	K =& \frac{1}{N-1} \left( \sum_{i, i\neq t}^{N} \sum_j^N W^{-1}_{ij} \right) \left( 1 + \frac{\kon'}{\koff'} \sum_n^{N_b} \sum_{m\in S} V^{-1}_{n m} \right).
      \end{split}
    \end{equation}
    Here $W$ is a matrix defining the spacial network, and $U$ ($V$) indicates the transition matrix on internal networks with (without) the target
    \begin{equation}\label{eq:transition}
        \begin{split}
        \koff' V &= L_\nu + \koff' \sum_{n\in S} E_{n n} ,\\
        \koff' U &= \koff' V + \Omega E_{t' t'} , \\
        \koff W  &= L_k + \koff I + \kon' \left( |S| - \sum_{n\in S}\sum_{m\in S} U ^{-1}_{n m} \right) E_{t t},
        \end{split}
    \end{equation}
    where $L_\nu$ and $L_k$ are Laplacian matrices of the internal and spatial networks, respectively. Here, $I$ represents the identity matrix, and $E_{n m}$ is a matrix with zero entries except for $(n,m)$, which is set to one. $|S|$ is the number of surface nodes (in general,  the size of the set $S$). Here, we considered a single surface node, i.e., $|S| = 1$.

    Combining Eqs.~\ref{eq:gen_total_c} and \ref{eq:gen_ct_cdna}, we obtain 
    \begin{equation} \label{eq:ctarget}
      \ctargetbar = \frac{\Gamma \bar c}{1 + K\frac{\kon}{\koff}(N-1)},
    \end{equation}
    which gives
    \begin{equation} \label{eq:J}
      \frac{J_\infty}{\overline{c}}  = \frac{\Omega\Gamma}{1 + K\frac{\kon}{\koff}(N-1)}
    \end{equation}
    from Eq. \eqref{eq:J}. This is our proxy for inverse MFPT and constitutes one of our manuscript's key results.

    If we consider a system lacking spatial or internal networks, Eq.~\eqref{eq:J} simplifies to
    \begin{equation}
     \frac{\Jinf}{\overline{c}} = \frac{\koff}{\koff + (N-1)\kon} \frac{\kon}{\Omega +\koff} \Omega.
    \end{equation}
    In other words, here, the bulk is in steady-state with $N$ nodes, where one contains a target that gets detected with rate $\Omega$. In our notation, this case is $K=1$ and $\Gamma = \kon /(\Omega + \koff)$.

\subsection{Optimizing flux: influence of rate constants in simple networks-within-network systems}
    
    This section explores how the rate constants impact the flux $\Jinf$ for two fixed internal network structures; a linear chain and a random Erd\H{o}s-R\'enyi network (ER). In particular, we are interested in parameter ranges that maximize the flux or minimize the search time. To keep the analysis tractable, we pay special attention to a ``free bulk''. This is a fully connected spatial network with the same on- and off-rates for all edges.
    In this case, $\Gamma$ and $K$ in Eq.~\eqref{eq:factors} simplify to
    \begin{equation}\label{eq:Factors_oneS}
        \begin{split}
    	\Gamma =&  \frac{\kon'}{\koff'}  \frac{ \kon U^{-1}_{t' 1}  }{ \koff + \kon' \left( 1 -  U^{-1}_{11} \right) },\\
    	K =&  1 + \frac{\kon'}{\koff'} \sum_n^{N_b}  V^{-1}_{n 1},
        \end{split}
    \end{equation}
    where $U^{-1}$ and $V^{-1}$ indicate the inverse of transition matrices $U$ and $V$, respectively. The absorbing rate $\Omega$ is in $\Gamma$ (via $U^{-1}$).

    \subsubsection{Effect of the off-rate from spatial network to bulk}
    
        First, we calculate how $\Jinf$ depends on the residence time associated with each node in the spatial network $1/\koff$. This parameter interpolates between two different search regimes, both having long search times. If $\koff$ is large, the searcher switches nodes rapidly and can scan a large part of the network. However, changing too frequently risks missing the target, even if being at the correct target node. In protein target search, this parameter represents the strength of unspecific binding between proteins and DNA.
    
        We illustrate this behavior in Fig.~\ref{fig:optimal_search} (two upper rows), where we plot the steady-state flux $\Jinf$ for varying $\koff$ using Eqs.~\eqref{eq:J} and \eqref{eq:Factors_oneS}. The solid lines represent analytical expressions, and the filled symbols show results from stochastic Gillespie simulations (see Methods). Regardless of the internal network structure, the curves follow the same general inverted parabola with a clear maximum for free bulk.
        The bottom row shows when the spatial network has a random ER structure. We note significant deviations for small $\koff$, where the flux $\Jinf$ is almost flat. It remains flat because the searcher jumps between the nodes in spatial networks, not through the bulk.

        From our general equation for $\Jinf$ (Eq.~\eqref{eq:J}), we may calculate analytically the maximum flux, $J_\infty^\textrm{max} = \Jinf(\kopti)$ for the ``free bulk'' cases. To this end, we compute $\partial \Jinf/\partial \koff = 0$ in and solve for $\kopti$. This gives
        \begin{equation} \label{eq:koff_optimal}
            \kopti = \sqrt{(N-1) K \kon \kon' \left(1-U^{-1}_{11}\right)}.
        \end{equation}
        Since $N$ and $K$ are network-specific parameters, this expression is summarized as
        \begin{equation} 
            \kopti \sim \sqrt{\kon \kon'}.
        \end{equation}
        In other words, the optimal off-rate is a geometric average between the on-rate from the bulk to the spatial node ($\kon$) and the on-rate from the spatial node to the internal network ($\kon'$). We show the analytically optimal points $\kopti$ and $\Jinf(\kopti)$ as black-filled circles in Fig.~\ref{fig:optimal_search}.
    
        Our exact expression for $\Jinf$ also allows us to understand better the small and large $\koff$ behavior of the flux. For small $\koff$, we find
        \begin{equation}
            \Jinf \simeq \frac{\Gamma}{K} \frac{\koff}{(N-1)\kon} \propto \koff.
        \end{equation}
        In this limit, the searcher gets stuck on non-target nodes in the spatial network. Increasing $\koff$ means making the searcher more mobile from the non-target nodes and results in an increased target flux $\Jinf$.
    
        In the other limit, $\koff$ is so large that the searcher switches spatial nodes too fast. It might even visit the target node several times without having time to cycle through the internal degrees of freedom before leaving. Expanding $\Jinf$ for large $\koff$ yields
        \begin{equation}
            \Jinf \simeq \Gamma -\Gamma K(N-1)\frac{\kon}{\koff} \propto \frac{\kon}{\koff} \frac{\kon'}{\koff'},
        \end{equation}
        where we used that $\Gamma \propto 1/\koff$ in Eq.~\eqref{eq:Factors_oneS} for large $\koff$ regime. This results in a leading term $1/\koff$.

        Another way to understand the $\koff$ behavior is to study the relative magnitudes of $\Gamma$ and $K$. They represent to what extent searchers interact with the spatial and internal networks, respectively. In Fig.~\ref{fig:optimal_search}, we note that $K\gg \Gamma$ for large $\koff$ indicates that the flux through the target is dominated by the internal degrees of freedom. Admittedly, this result is relatively trivial in the two first examples studied in Fig.~\ref{fig:optimal_search} because the spatial network is so simple (fully connected). 
        
        However, with a spatial network, the values of $\Gamma$ and $K$ are less trivial. Here, $\Gamma$ decreases monotonically, whereas $K$ increases until a plateau. When $\koff$ is small, we note that the spatial network dominates the diffusive dynamics, effectively reducing the effect of the non-target nodes. Studying $\Gamma$ and $K$ is a general approach to understanding where the searcher spends the most time: on the spatial network or cycling through the internal degrees of freedom.

    \subsubsection{Effect of on-rate from spatial to internal network}
    
        \begin{figure}
            \centering
            \includegraphics[width=\columnwidth]{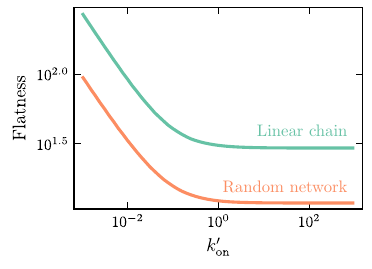}
            \caption{
            Inverse curvature (flatness) of $\Jinf$ versus $\kon'$ calculated from the analytical curves in Fig.~\ref{fig:optimal_search}. Here, we show the two cases of the internal networks: the linear chain and the random internal network.
            The width of the plateau increases for smaller $\kon'$, which saturates at larger values.
            On- and off-rates for all edges in the internal networks are the same.
            }\label{fig:konp_flatness}
        \end{figure}
        
         Next, we study how the target flux changes with the rate of entering the internal degrees of freedom $\kon'$. In particular, we note that a plateau emerges in $\Jinf$ around $\koff \approx \kopti$ in Fig.~\ref{fig:optimal_search}, where it expands as $\kon'$ decreases.
         We interpret this as the $\Jinf$ cannot increase beyond its maximum value $\Jinf^\textrm{max}$, and further increase in $\koff$ does not enhance $\Jinf$ due to the limiting effect of $\kon'$, with the relationship $\koff^*\propto\sqrt{\kon'}$.
         Eventually, as $\koff$ exceeds $\koff^*$, $\Jinf$ starts to decline as there is a significant chance of missing the target.
         In this limit, $\Jinf \propto \kon'/\koff$, as derived above.
        
        Just as there is an optimal off-rate $\kopti$, there is an optimal $\kon'^*$ that maximizes the flux, $J_\infty^\textrm{max} = \Jinf(\kon'^*)$. Performing a similar analysis as before, by solving $\partial \Jinf/ \partial\kon' = 0$, leads to   
        \begin{equation} \label{eq:konp_optimal}%
            \kon'^* = \sqrt{\frac{\koff'\koff\left[1 + \frac \koff {(N-1)\kon}\right]}{\left(1-U^{-1}_{11}\right)\sum_n^{N_b}V^{-1}_{n1}}}.
        \end{equation}

        This equation has two distinct behaviors for small and large $\koff$. When large, $\koff$ and $\kon'$ are proportional to each other, $\kon'^*\sim \koff$. But when $\koff$ is small, they follow a sublinear relationship, $\kon'^*\sim \sqrt \koff$.

        Next, we study how the plateau in $\Jinf$ develops for growing $\kon'$. To this end, we calculate the curvature and plot its inverse, or ``flatness'' $F$:
        \begin{equation}
             \frac 1 F = \left. \frac{\partial^2\Jinf(\koff)}{\partial \koff^2}\right|_{\koff=\kopti}.
        \end{equation}
        As shown in Fig.~\ref{fig:konp_flatness}, $\Jinf$ exhibits high flatness at small $\kon'$; however, as $\kon'$ increases, it becomes easier to enter the internal network, and the flatness decreases with a more pronounced peak in $\Jinf^\textrm{max}$.
        At some $\kon'$, the peak is not getting sharper, and the flatness settles at a low value.

        This overall decrease in $F$ with $\kon'$ does not change in different internal network structures, but we noted a slight shift.
        Specifically, for a linear chain, a slight deviation of $\koff$ from $\koff^*$ results in smaller change to $\Jinf$ from its optimum $\Jinf^\textrm{max}$ compared to a random network, as shown in Fig.~\ref{fig:konp_flatness}.
        However, the flux $\Jinf$ in the linear chain is lower, due to its sequential structure: once trapped inside, a searcher has to travel through to surface node from internal nodes one by one to exit.
        In contrast, the surface node in random network has multiple connections to other internal nodes, allowing for easier exits via shortcut edges.
        This structural difference also affects the response of $F$ to the changes in $\koff$ and $\kon'$.
        The single connection to the internal node from the surface node in linear chain limits the response of $F$ to changes in $\koff$ and $\kon'$.
        In contrast to the linear chain, several shortcuts to surface node in random network allow the stronger deviation in $\Jinf$, which results in lower $F$, the sharper peaks on both rates $\koff$ and $\kon'$, as shown in Figs.~\ref{fig:optimal_search} and \ref{fig:konp_flatness}.

    \subsubsection{Effect of the absorption rate}

    \begin{figure}
        \centering
        \includegraphics[width=\columnwidth]{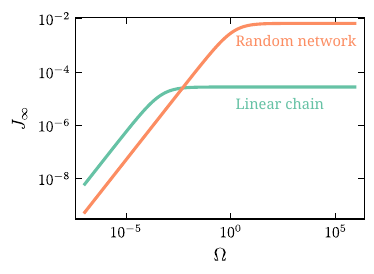}
        \caption{The flux $\Jinf$ versus the target absorption rate $\Omega$ for a skewed internal network structure where $\nu_{nm} = 2\nu_{mn}$ if $n > m$. For large $\Omega$, the flux saturates as the target node $c_t^{t'}$ becomes perfectly absorbing. However, for some threshold of $\Omega$ the linear chain network has a larger flux than the random network.}
        \label{fig:J_omega}
    \end{figure}

    In our model, $\Omega$ represents the absorption rate from exiting the system once the searcher reaches the target node and attains the correct internal state. In cell biology, this could mean the rate at which a desired multi-subunit complex gets formed that is able to repress or enhance a gene's activity. In the cases studied above, we considered a relatively large $\Omega$, making it not rate-limiting. Here, we extend our analysis to varying $\Omega$ in different internal network structures.

    To this end, we constructed an asymmetric internal network, where the rate of going deeper into it is twice as large as the rate of leaving, i.e., $\nu_{nm} = 2\nu_{mn}$, for $n > m$. Doing this for the linear chain and the random network, we get the results in Fig.~\ref{fig:J_omega}. As expected, the flux saturates for $\Omega \rightarrow \infty$, as the absorption rate of the target is no longer rate limiting.
    We also note that the flux $\Jinf$ increases for small $\Omega$ values. In our particular cases,  $\Jinf\propto \Omega$. These behaviors are similar for both two internal network structures.

    However, there are quantitative systematic differences. The random network shows a larger flux for large $\Omega$, as the network has shorter mean path lengths $D$ in the internal network. However, before some threshold value, this is no longer true, and the linear chain predicts a larger flux than the random network for smaller $\Omega$. We can interpret this based on the network structure. The random network has more links to the target, but that also has more links away from the target. If the absorption rate $\Omega$ is too low --- the searcher might just leave the target node without absorbing. The same might happen for the linear chain, but since it only has one rate of leaving, this chance is reduced and leads to an overall increase in the flux.
    Furthermore, the only edge has asymmetric rates; the rate of going out is smaller than the rate of going deeper down.
    As such, the linear chain network might be a practical internal structure that `stabilizes' slow reactions at the target in the small $\Omega$ regime, overall decreasing the search time.

    \subsection{Optimal fluxes and network sizes}

        \begin{figure}
        \centering
        \includegraphics[width=\columnwidth]{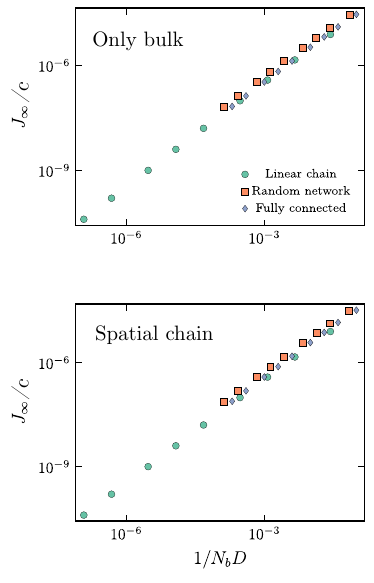}
        \caption{Flux $\Jinf$ versus inverse of internal system size $N_b$ and mean shortest path length $D$ for \textbf{(upper)} free bulk space, and \textbf{(lower)} linear chain spatial network with $N=1000$ nodes.
        The flux increases in $1/N_b D$ linearly.
        }
        \label{fig:J_NbD}
        \end{figure}

        \begin{figure}
        \centering
        \includegraphics[width=\columnwidth]{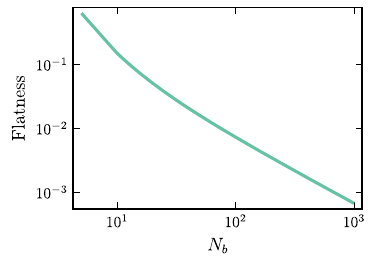}
        \caption{The inverse curvature (flatness) as in Fig.~\ref{fig:konp_flatness} against internal network size $N_b$ with the linear chain. Barring an initial faster decrease, the plateau decreases as a power law relative to the internal network size.}
        \label{fig:Nb_flatness}
        \end{figure}
        
        We quantify the system size by the number of nodes $N$ for the spatial network and $N_b$ for the internal network. These are key parameters influencing flux $\Jinf$. Intuitively, it becomes more challenging for a searcher to find the target as the network size increases, leading to a smaller $\Jinf$.
        For example, in a system with a fully connected spatial network (``free bulk'') with uniform on- and off-rates, the searcher must find a target among $N$ possible sites. This requirement imposes a scaling behavior of $\Jinf \sim N^{-1}$.
        As shown in Eq.~\eqref{eq:J} and in Fig.~\ref{fig:optimal_search} (middle column), $\Jinf$ decreases with as $\Jinf \propto 1/[1+ \textrm{const.}\times (N-1)]$, that becomes $\Jinf \sim N^{-1}$ for large $N$.

        The internal network structure is also essential for $\Jinf$.
        For fully-connected or random internal networks, $\Jinf$ decreases as $\Jinf \sim N_b^{-1}$.
        However, this contrasts a linear chain that shifts this relationship to $\Jinf \sim N_b^{-2}$.
        This difference originates from the structural properties of the network.
        The contribution to $\Jinf$ due solely to network size $N_b$ follows a scaling of $\sim N_b^{-1}$, but this is compounded by the network’s structural influence, which relates to the mean shortest path length $D$.

        This length changes in system size~\cite{barabasi2013network}.
        For example, in a fully connected network, $D$ remains constant, $D=1$, regardless of the network size. However, in a random network, it grows with $N_b$ as $\ln N_b$.
        Therefore, by incorporating the system size and the structural factor, through $D$, for both fully connected and random networks, the leading term in $\Jinf$ scales as $N_b^{-1}$.
        On the other hand, for a linear chain, the $D$ scales linearly with $N_b$, resulting in a slower scaling of $\Jinf \sim N_b^{-2}$.
        
        Figure~\ref{fig:J_NbD} shows how the flux $\Jinf$ changes with the internal system size $N_b$, renormalized by mean path length $D$ of the internal network for free bulk and a linear spatial network with $N=1000$ nodes. We find that the curves agree with $\Jinf \propto 1/N_b D$ in all cases.

        We also study how the flux $\Jinf$ behaves near the optimal flux $\Jinf^{\textrm{max}}$ as a function of $N_b$, by calculating the flatness $F$.
        Figure~\ref{fig:Nb_flatness} shows the decrease in  $F$ for the internal linear chain $N_b$.
        We note that as $N_b$ increases, $\Jinf$ decreases, and the change in $\Jinf$ is larger for small variations in $\koff$. This means that a more internal degrees of freedom leads to generally slower search time and smaller robustness near the optimal flux.

\subsection{Impact and robustness of internal network structure on target flux}\label{subsection:robustness}

    \begin{figure*}
        \centering
        \includegraphics[width=\textwidth]{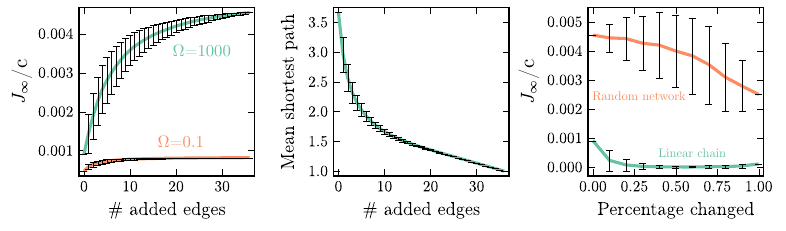}
        \caption{How flux robustness depends on changes in link density and reaction rates.
        \textbf{(left)} To see how the flux $\Jinf$ changes when the network switches between the linear chain $\rightarrow$ random and $\rightarrow$ fully-connected networks, we started with a linear chain and added links randomly. The line shows the analytical calculation, with standard deviations between realizations as error bars. The flux increases with more edges, but only at larger absorption rates $\Omega$.
        \textbf{(middle)} The increase in flux can be seen as a change in the mean shortest path length of the network, which we calculated from the data in the left figure. As more edges are added, the shortest path decreases.
        \textbf{(right)} We analytically calculated the flux $\Jinf$ versus a percentage of links set to a tenth of the original rate. Error bars show the standard deviation between realizations. In the random internal network, mutations slow down the search. However, this is only true for the linear chain up to some point, where the flux increases with more mutations.
        }
        \label{fig:robustness}
    \end{figure*}

    The previous sections studied target fluxes $\Jinf$ for varying dynamical parameters such as $\koff$, $\kon'$, and $\Omega$ in a few fixed network structures and general system sizes ($N$ and $N_b$). This section focuses on the internal network connectivity. In particular, how the target flux changes with varying network connectivity and link weights.
    
    First, to establish a base case, we consider the linear chain where the ratio, or ``quotient'' $C$, of all on/off rates are identical, i.e., $C=\kon'/\koff' = \nu_{nm}/\nu_{mn}$. This case gives a relatively simple expression for $K$:
    \begin{equation}\label{eq:linearK}
    K = \sum_{l=0}^{N_b} C^l = \frac{1 - C^{N_b + 1}}{1 - C}.
    \end{equation}

    This expression has interesting limits with respect to the flux $\Jinf$. If we consider a free bulk spatial network, then $\Jinf \propto 1/ (1+\textrm{const.}\times K$). When $C$ is small, such as when the exit rate $\koff'$ is large, the searcher spends little time in the internal network and $K$ is relatively moderate $K \approx 1+C$ ($C \ll 1$). Therefore, in this limit marginally affects $\Jinf$. However, things change when the quotient is $C\geq 1$. Here, we find $K\approx C^{N_b}$ when $C\gg 1$ and $\lim_{C\to 1}K=N_b+1$. This means that $\Jinf$ decreases a power law with the exponent $N_b$ when $C>1$, and the searcher will have difficulty finding the target for large $N_b$.
    We present the detailed derivation of factors for the linear chain in Appendix~\ref{appx:linear}.
    
    Next, we consider a random network configuration (ER). In this case, a searcher can more easily escape exploring the spatial network and bulk, thus, in general, reducing search time if $\Omega$ is large. We studied this case numerically, starting with a linear chain with a fixed quotient $C$, and then randomly adding new the random network symmetrically. By adding new edges, the internal network goes through three phases: linear chain, random network (ER), and fully connected. For each network configuration, we calculate $\Jinf$ as shown in Fig.~\ref{fig:robustness}. As expected, adding new links increases the flux into the target, especially when the target absorption $\Omega$ is large. However, when $\Omega$ is small, this increase is no longer trivial. Instead, we find there exists an optimal number of links that maximizes the target flux. Like before, there is a trade-off where quickly searching and escaping the internal network is fighting against the risk of missing the target.

    For completeness, we plot how the mean path length $D$ of the internal network shrinks with increasing link density in Fig.~\ref{fig:robustness}. This figure shows a steep decline that stops when all the internal distances are equal to one.

    In addition to adding new links, we also investigate how $\Jinf$ changes with varying rates between the internal degrees of freedom (see right figure in Fig.~\ref{fig:robustness}). To this end, we start with a random network and reduced a fixed fraction of the links by a factor of 10. This approach mimics mutations to enzymes in biochemical reaction networks ~\cite{lind2019predicting}. We find that increasing the percentage of changed links not only reduces the target flux $\Jinf$ but also increases the variability (see error bars). Because of this large variability, it is difficult to define a typical target-finding time.

    Next, we conduct similar link-weight alterations for linear chains with constant quotients $C$. In contrast to the random network case, the linear chain exhibits a significantly steeper initial decrease. Additionally, there is a slight increase towards the right end of the curve. We interpret this as follows: if only a few of the rates in the linear chain are altered, the searcher may become trapped within ``wrong'' internal networks, as there are no alternative pathways around bottlenecks. However, as the number of alterations increases, these bottlenecks disappear, resulting in increased flux $\Jinf$.

\subsection{Application: DNA-target search}

    Throughout the paper, we've discussed our network-within-networks model from general problems in complex systems. However, our framework adds specific value for the general understanding of site-specific search by DNA-binding proteins, most notably Transcription Factors (TF),  studied by several research groups~\cite{hedstrom2023modelling,klein2020skipping,bauer2012generalized,benichou2011intermittent,slutsky2004kinetics}.

    The most common form of gene regulation occurs when a TF protein binds to a short DNA sequence near the start of a gene, either initiating or repressing transcription. These regulatory sequences are very short relative to the total length of DNA ($\sim 10^{-8}$ in humans), making it a challenging task for TFs to find the correct binding sequence, especially when there are many similar "almost targets" scattered across the genome. If the TF is highly sensitive to the exact sequence, it will bind successfully upon encountering the target. However, this sensitivity can slow down the search because the TF may get trapped at incorrect sites. Reducing sensitivity can speed up the search, but if reduced too much, it also lowers the chance of binding to the target, even when the TF is directly over it. This trade-off is known as the speed-stability paradox\cite{benichou2009searching}. One resolution to this paradox is the introduction of at least two internal states for the TF: a fast ``search mode'' that allows rapid movement across large genomic regions and a slow ``recognition mode'' to ensure binding once the correct target is found. One resolution to this paradox is introducing internal states. For TFs, these states represent a fast search mode, allowing the TF to move over large genomic regions without, and a slow recognition mode with target detection. 

    Our theory allows us to study optimal parameter choices associated with TF search. For example, the sliding length $\ell$. This is the DNA-segment length the TF diffuses before detaching. If $D_\textrm{DNA}$ is the TF's diffusion constant, then $\ell =(2D_\textrm{DNA}/\koff)^{1/2}$. Using the optimal $\koff$ from our theory, we are able to estimate the optimal sliding length as $\ell^*=(2D_\textrm{DNA}/\kopti)^{1/2}$. From Eq.~\eqref{eq:koff_optimal}, we find that the optimal off-rate is proportional to a simple geometric average between the rates of switching to recognition mode $\kon'$ and search mode $\kon$. This gives the optimal sliding length $\ell^*\sim (2D_\textrm{DNA})^{1/2} (\kon \kon')^{-1/4}$. Another way to interpret $\kon$ is the binding rate to DNA from the surrounding bulk. Therefore, in this case, $\kon$ is the rate per DNA segment, where $N\kon$ is the total binding rate to DNA; $N\gg 1$ is proportional to the DNA length. Finally, we may also generalize the optimal sliding length to include more than two TF states. This is captured in $K$, where, for example $K \approx 1 + \kon'/\koff' + (\kon'/\koff')\times(\nu_{21}/\nu_{12}) + \cdots$ for the linear chain (see Appendix~\ref{appx:linear}). Putting it all together, we find a novel scaling relationship for the sliding length that minimizes search times for DNA targets
    \begin{equation}
        \ell^* \sim D_\textrm{DNA}^{1/2} (N\kon \kon')^{-1/4} K^{-1/4}.
    \end{equation}

    In addition to studying optimal dynamical rate constants, we analyzed how search times are influenced by rewiring the network of internal states. For DNA-binding proteins, these states can represent conformational changes, phosphorylation, binding of small molecules, or interactions with general transcription factors. Rewiring or changes in network weights could correspond to genetic mutations or environmental influences.

    Having a large internal network ($N_b \gg1 $) may increase the search time, but it is more robust to weight alterations (see Fig.~\ref{fig:J_NbD}), such as those caused by mutations. Notably, this robustness depends on the target absorption rate $\Omega$. If the target absorption rate is too low, a large internal network can lead to a slowdown, as additional links may increase the risk of missing the target. In such cases, a linear chain of reactions can help stabilize target-finding times. These structural effects highlight the trade-off between stability and fast target finding. This ties back to our earlier discussion on how changes in internal states or network weights, such as those caused by genetic mutations or environmental factors, can affect search efficiency.
    
    \section{Conclusion}
    In real-world applications, search processes typically occur across layers of weakly connected networks.
    Examples range from large-scale to small-scale systems, such as finding email spammers inside local area networks or target-search processes in cells.
    This paper lays out a theoretical framework for these types of processes.
    By considering the input from bulk, random diffusion on the spatial network, and the diffusion in the internal networks, one of which contains the target, we have studied how the target search times change in two different networks, spatial and internal networks.
    Our theory allowed us to derive universal expressions that hold for a broad class of networks-within-network organizations and showed excellent agreement with simulations.

    Foremost, we have studied target search time, which is represented in this paper as the inverse of the target flux $\Jinf$, and provided a general understanding of the critical variables that have a significant impact on the search.
    We have obtained the general expression of the $\Jinf$ in terms of two factors $\Gamma$ and $K$, which contain the information of structure of networks-within-network, and analyzed the optimal flux $\Jinf^{\textrm{max}}$ by varying the off-rates $\koff$ to the bulk.
    We also calculated how strongly the flux $\Jinf$ deviates from the optimum from small perturbations of the rates related to the internal structure.

    In terms of the absorbing rate $\Omega$, for large $\Omega$ regime, $\Jinf$ does not change in both the linear chain and random network, and $\Jinf$ in the random network is higher than in the linear chain, which is related to the higher connectivity of the random network.
    In the small $\Omega$ regime, however, even $\Jinf$ increases linearly for both internal networks, linear chain has higher $\Jinf$.
    
    We have also shown how the size and structure of networks-within-networks, particularly the spatial network size $N$ and the internal network size $N_b$, significantly influence the target search flux $\Jinf$. As the spatial network grows bigger, $\Jinf$ generally decreases, indicating increased search difficulty.
    Notably, the internal network structure also plays a role, where $\Jinf$ is proportional to the inverse of system size $N_b$ and mean path length $D$ of the internal network.

    Furthermore, we have also investigated how the internal network structure and the internal rates change the target search time.
    For structural effects, we observed changes in the flux $\Jinf$ by adding new edges to the linear chain.
    The edge-adding process results in a change in the internal network structure from a linear chain to a fully connected network, and it definitely affects search time by helping the searcher not to get stuck within the internal network.

    We also applied our theory to DNA-target search by Transcription Factors (TFs). Another example from cell biology that could be analyzed using our framework is RNA polymerases (RNAp). RNAps are multi-subunit protein complexes, typically much larger than TFs, that bind to promoter regions to read the gene and synthesize the RNA strand, serving as the template for protein synthesis. To start transcription, the RNAp passes through several internal initiation steps that are similar to a linear chain of states (e.g., see McClure's 3-step model for transcription initiation)~\cite{mcclure1980rate}. In addition, cells use a broad collection of proteins and small molecules that bind to RNAps to alter their internal state, such as sigma factors. These factors are often part of stress-response systems and help RNAp recognize and transcribe select sets of promoters. These internal states do not necessarily follow a causal order and could be represented as a random internal network.

    Here, we studied the diffusion on networks-within-network using three transition matrices.
    Notably, our model can be extended to the random diffusion on general multiplex networks that include both source and sink.
    In this extension, the random diffusion dynamics is governed by a supra-transition matrix.
    This supra-transition matrix comprises a large block-diagonal matrix of Laplacian matrices, representing the diffusion on each network layer, along with additional matrices that capture interactions between different layers as well as the source and sink.

    To close, this paper offers a new framework to better understand search and navigation processes in complex systems that often can be represented as a hierarchical networks-within-network structure.
    We believe that our specific case of multiplex network structure can be applied to many disciplines and that our findings can be used to better understand a range of physical and biological systems.
  
    \begin{acknowledgements}
    We want to thank Yuri Schwartz at Umeå University for valuable discussions.
    We acknowledge financial support from the Swedish Research Council (grant no. 2017-03848).
    S.G.Y was supported by a postdoctoral fellowship from the Carl Tryggers Stiftelse.
    The computations were enabled by resources provided by the National Academic Infrastructure for Supercomputing in Sweden (NAISS) at High Performance Computing Center North (HPC2N), partially funded by the Swedish Research Council through grant agreement no. 2022-06725.
    \end{acknowledgements}

    \appendix
\section{Derivation of $\Gamma$ and $K$ for General Cases} \label{appx:derive}
	Utilizing the transition matrices $W$, $V$, and $U$ in Eq.~\eqref{eq:transition}, we obtain the steady-state solutions of Eq.~\eqref{eq:EoM} as
	\begin{equation} \label{eq:steady_state_cs}
	\begin{split}
	&\bar c_i = \cbulkbar\frac{\kon}{\koff} \sum_j^{N} W^{-1}_{i,j},\\
	&\bar c^n_i =  \bar c_i \frac{\kon'}{\koff'}\sum_{m\in S}V^{-1}_{n,m},\\
	&\bar c_t^n = \bar c_t \frac{\kon'}{\koff'}\sum_{m\in S}U^{-1}_{n,m}.
	\end{split}
	\end{equation}
    Combining this steady-state solutions, Eqs.~\eqref{eq:gen_J},~\eqref{eq:gen_total_c}, and \eqref{eq:gen_ct_cdna}, we get the factors $\Gamma$ and $K$ for the general case
	\begin{equation}
	\begin{split}
	\Gamma =& \frac{\kon}{\koff}\frac{\kon'}{\koff'} \left( \sum_j^N W^{-1}_{t,j} \right) \left( \sum_{m\in S} U^{-1}_{t', m} \right),\\
	K =& \frac{1}{N-1} \left( \sum_{i, i\neq t}^{N} \sum_j^N W^{-1}_{ij} \right) \left( 1 + \frac{\kon'}{\koff'} \sum_n^{N_b} \sum_{m\in S} V^{-1}_{n,m} \right).
	\end{split}
	\end{equation}
    In $\Gamma$, the absorbing rate $\Omega$ is included inside the inverse matrix term $\sum_{m\in S}U^{-1}_{t'm}$.
	For large spatial network, the factor $K$ is approximately $K \approx 1 + \frac{\kon'}{\koff'}\sum_{n}^{N_b} \sum_{m\in S} V^{-1}_{n, m}$, because we can ignore the contribution of the spatial target node on $K$.
	In the case where all the internal networks are different for every nodes in space network, the results are almost the same, but the transition matrix $V$ depends on the node $i$ in the space network as $V^{i}$.
	Thus, the summation over $i$ in $K$ should be applied thoroughly.
    For the free bulk system, we can rewrite the $\Gamma$ as $\Gamma = \frac{\kon}{\koff}\frac{\kon'}{\koff'} \left( \sum_{m\in S} U^{-1}_{t', m} \right)$, and $K$ is the same as that in large $N$ limit.
    When only single internal node has the connection to spatial network, $\Gamma$ and $K$ give Eq.~\eqref{eq:Factors_oneS} in main text.

 \section{Free Bulk and Linear Chain Internal Network} \label{appx:linear}
    
    Nodes are connected in a row in the linear chain, and the nodes have links when $n = m-1$ and $n = m+1$.
    Diffusion dynamics in Eq.~\ref{eq:EoM} on internal linear chain can be rewritten as  
    \begin{equation}
    \begin{split}
      &\dv{c_i}{t} =  \kon \cbulk  - \left(\koff + \kon'\right) c_i + \koff' c_i^1,\\
      &\dv{c_i^1}{t} =  \kon' c_i  - \left(\koff' + \nu_{21}\right) c_i^1 + \nu_{12} c_i^2,\\
      &\quad \quad \vdots\\
      &\dv{c_i^n}{t} =  \nu_{n,n-1} c_i^{n-1}  - \left(\nu_{n-1,n} + \nu_{n+1,n}\right) c_i^n + \nu_{n,n+1} c_i^{n+1},\\
      &\quad \quad \vdots\\
      &\dv{c_i^{N_b}}{t} =  \nu_{N_b,N_b-1} c_i^{N_b-1}  - \left(\nu_{N_b-1,N_b} + \Omega \delta_{it} \right) c_i^{N_b} .
    \end{split}
    \end{equation}
    %
    In steady state, $\ctargetbar = \bar c_t^{N_b}$ becomes
    \begin{equation}
      \bar c_t^{N_b} = \frac{\nu_{N_b,N_b-1}}{\nu_{N_b-1,N_b} + \Omega} \bar c_t^{N_b-1}.
    \end{equation}
    
    Considering that the first internal node $c_i^1(t)$ is connected to $c_i(t)$ via $\kon'$ and $\koff'$, we can calculate $K$ as
    \begin{equation}
    \begin{split}
    K &= 1 + \frac{\kon'}{\koff'} + \frac{\kon'}{\koff'}\frac{\nu_{21}}{\nu_{12}} + \frac{\kon'}{\koff'}\frac{\nu_{21}}{\nu_{12}}\frac{\nu_{32}}{\nu_{23}} + \cdots \\
      &= 1 + \frac{\kon'}{\koff'}\left(1 + \sum_{l=2}^{\Nstates}\prod_{n=1}^{l-1} \frac{\nu_{n+1,n}}{\nu_{n,n+1}}\right).
    \end{split}
    \end{equation}
    In the simplest case where we consider only one node in the chain and the target to be at the first node in the internal network, $K = 1 + \kon'/\koff'$.
    When all the on/off rate quotient $C = \nu_{n+1,n}/\nu_{n,n+1}$ are the same for $\forall n$, and $C=\kon'/\koff'$, it gives the $K$ as shown in Eq.~\eqref{eq:linearK}.

    When the target is at the bottom at the linear chain, i.e., $\bar c_t^{N_b} = \bar c_t^{t'}$ the concentration at the target internal node becomes
    \begin{equation}
      \bar c_t^{N_b} = \prod_{n=0}^{N_b} \Omega_n \cbulkbar,
    \end{equation}
    where
    \begin{equation}
    \begin{split}
      \Omega_0 &= \frac{\kon}{\koff + \kon' - \koff'\Omega_1},\\
      \Omega_1 &= \frac{\kon'}{\koff' + \nu_{2, 1} - \nu_{1, 2}\Omega_{2}},\\
      \vdots\\
      \Omega_n &= \frac{\nu_{n, n-1}}{\nu_{n-1, n} + \nu_{n+1, n} - \nu_{n, n+1}\Omega_{n+1}},\\
      \vdots\\
      \Omega_{N_b} &= \frac{\nu_{N_b,N_b-1}}{\nu_{N_b-1,N_b} + \Omega},
    \end{split}
    \end{equation}
    which gives
    \begin{equation} 
      \Gamma = \prod_{n=0}^{N_b} \Omega_{n}.
    \end{equation}
    For the simplest chain, $\Gamma = \kon/(\koff + \Omega)$.

    We calculated the flux using a constant quotient $C$, shown in Sec.~\ref{subsection:robustness}. If each quotient is equal, there is a value that maximizes the target flux, in this case  $C = 1$. This value balances the return rate at each non-target site while maintaining the possibility of reaching the bottom of the internal network once at the target site.

\section{Optimal quotients in the linear chain internal network}

    \begin{figure*}
        \centering
        \includegraphics[width=\textwidth]{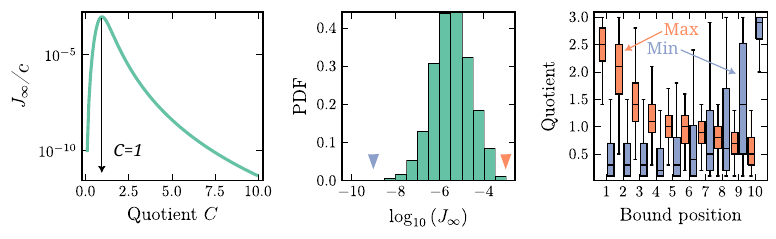}
        \caption{Understanding how the quotient between on/off-rates affect the linear chain internal network. Here we only consider bulk (no spatial network).
        \textbf{(left)} The search flux $\Jinf$ versus a constant quotient $C$. The flux increases up to a maximum at $C=1$, to then decrease for bigger quotients.
        \textbf{(middle)} We randomly sampled the quotients in the linear chain and plotted the flux in a histogram. At the ends we have a set of quotients in the linear chain that maximizes and minimizes the flux, shown by the orange and blue arrows.
        \textbf{(right)} We retrieved the sets of quotients that minimized and maximized the flux in the middle histogram for many samples and plotted their values using box plots. The orange (blue) boxes show the set of quotients that maximize (minimize) the flux. The optimal set of quotients in this case increases the flux compared to the fix quotient of $C=1$ by 40\%.
        }
        \label{fig:lc_quotients}
    \end{figure*}

    To better understand how the rates in the linear chain affects the search flux we analytically calculated the flux for different cases; one with constant quotients $C$ and with randomly sampled quotients $U(0, 3)$. These results are shown in Fig.~\ref{fig:lc_quotients}. Using a fix quotient, we note that there is a clear maximum in the flux for a specific value, in this case $C=1$. For low quotients, the system is stuck on the wrong sites, and never reaches the target. If the quotients are large, the searcher might never find the correct internal node. Note however that this curve is not symmetric, and the flux is less sensitive to increases in the quotients, rather than increases. This means that a big chunk of the search time is restricted by time spent on other sites. This is logical, since there are more non-target than target sites.

    However, a fix quotient might not be the optimal configuration for this system. Rather, a set of different reaction rates might correlate as to create a larger flux. To investigate this, we sampled random sets of quotients by drawing random numbers from $U(0, 3)$ and plotted the flux in a histogram. These results are shown in Fig.~\ref{fig:lc_quotients} (middle). This distribution shows a log-normal behaviour due to the correlations between the quotients in the linear chain. From this histogram we can extract the set of quotients that maximizes and minimizes the flux, shown by the arrows.
    
    By calculating many histograms and storing the quotients, we can show the set that maximizes and minimizes the flux, seen in Fig.~\ref{fig:lc_quotients} (right). Interestingly enough, the maximum flux case here yields a 40\% speed increase compared to the fix quotient $C=1$ case, indicating that a heterogeneous set of reaction rates is beneficial. From the plot we can see that the maximum case allows for fast binding at the shallow nodes in the network. However, going deeper into the network is more difficult. This strategy allows for fast probing the internal networks whilst not getting too stuck in the deeper nodes. However, in the slow case the searcher rarely enters the network, but when it does, it has a high probability of getting stuck deep into the network and never leave. In summary, the searching strategy should investigate the network shallowly, and not spend too much time deep in the network.

    \bibliography{../refs}

\end{document}